\documentclass[twocolumn,10pt]{article}
\usepackage{geometry}
\usepackage{hyperref}
\usepackage[strings]{underscore}
\usepackage{graphicx}
\usepackage{amsmath}
\usepackage{url}
\usepackage{amsfonts}  
\usepackage{amssymb}

\title{\bfseries Cognitive Biases at Play? Insights from a Bayesian Game Framework}
\author{
  Samiha Tariq\\[0.5em]
  \ Southern Illinois University Carbondale
}
\date{May 2025}

\usepackage{abstract}

\begin{document}

\twocolumn[
\begin{@twocolumnfalse}
\maketitle
\begin{center}
\begin{minipage}{0.85\textwidth}
\begin{abstract}
\centering
\itshape

This paper examines the impact of cognitive biases on financial decision-making through a static Bayesian game framework. While traditional economic theories typically assume investor rationality, this research recognizes that real-world decisions are often influenced by cognitive biases such as loss aversion, overconfidence, and herd behavior. Integrating psychological insights with economic game theory, the study models strategic interactions among investors deciding between risky and risk-free assets. Utilizing Bayesian Nash Equilibrium, the analysis demonstrates how individual biases distinctly affect investment choices and market dynamics. Results align with Herbert Simon's theory of bounded rationality, indicating that cognitive biases can significantly diverge decision-making from rational economic assumptions, potentially leading to market inefficiencies, price bubbles, and financial crashes. This interdisciplinary approach emphasizes the necessity of incorporating psychological factors into economic models, providing valuable insights for policymakers aiming to foster greater market stability and more informed financial decision-making.

\end{abstract}
\end{minipage}
\end{center}
\end{@twocolumnfalse}
]

\section{Introduction}

This study investigates how cognitive biases shape financial decision-making through the lens of game theory, specifically using a static Bayesian game model of incomplete information. Traditional economic theories typically suggest that investors aim to maximize returns through rational decision-making. However, behavioral economics highlights that investor behavior can deviate from pure rationality due to cognitive and psychological factors. Simon's (1956) theory of bounded rationality explains that cognitive constraints, computational limits, and incomplete information can lead investors to suboptimal choices.

Cognitive biases—systematic deviations from rational judgment driven by personal experiences, preferences, and mental shortcuts—significantly influence investment decisions. Notably, Kahneman and Tversky's (1979) concept of loss aversion illustrates that individuals experience the pain of losses about twice as intensely as the pleasure from equivalent gains. This bias often results in excessive caution, causing investors to shy away from potentially lucrative opportunities. Similarly, herd behavior, the tendency of individuals to mimic others’ financial decisions instead of conducting independent analysis, contributes substantially to market volatility. For example, Dhungana et al. (2022) demonstrate how herding significantly affects investor behavior in Nepal’s stock market, highlighting its capacity to disrupt rational decision-making processes.

Moreover, overconfidence—characterized by an excessive belief in one's predictive abilities—is prevalent among investors. Research by Barber and Odean (2001) indicates that overconfident investors trade approximately 45\% more frequently than their less confident counterparts, resulting in lower returns and heightened market risks.

By integrating psychological insights into financial analysis, this research employs a static Bayesian game framework to model the strategic interactions shaped by cognitive biases such as risk aversion, overconfidence, and herd behavior. Unlike prior econometric studies by Nur Aini and Lutfi (2019) and Zhang et al. (2022), a game-theoretic approach introduces a novel perspective, emphasizing the interactive and strategic dimensions of investor decision-making. This innovative modeling helps illustrate how cognitive biases can trigger market anomalies, including bubbles, volatility, and financial crises.

Understanding these cognitive biases is crucial not only for theoretical clarity but also for practical policy formulation. Insights from this research can guide policymakers and financial institutions in developing strategies to mitigate bias-driven volatility, promote informed decision-making, and stabilize financial markets. Additionally, the findings offer potential for creating financial products tailored specifically to real-world investor behaviors and psychological characteristics, fostering more inclusive and resilient financial market practices.

\section{Literature Review}

Simon (1956) first challenged traditional economic theories of rationality, proposing the concept of "bounded rationality" to emphasize that rational choice is constrained by internal cognitive limitations and external environmental factors. He highlighted that human cognitive capabilities are inherently limited, making it impossible for individuals to process every piece of available information or anticipate all possible outcomes due to constraints in knowledge, attention, and computational capacity. Consequently, decision-makers rely on a simplified understanding of reality rather than exhaustive rational analysis, opting for choices that merely satisfy an acceptable level of satisfaction—a process Simon termed "satisficing."

Moreover, Simon identified asymmetric information as a critical factor constraining rationality, wherein information is unevenly distributed, causing decision-makers to lack complete knowledge about their circumstances and the potential consequences of their actions. This informational imbalance further distorts decision-making processes, leading to less optimal outcomes. Simon emphasized that individuals form perceptions and make decisions within their neuro-environment, shaped significantly by available information and inherent cognitive biases, underscoring the necessity of incorporating psychological dimensions into economic rationality theories.

Building upon Simon's theoretical framework, recent literature has extensively explored the impacts of cognitive biases on investment decisions. Zik-Rullahi et al. (2023), through secondary data analysis, corroborate Simon's notion by concluding that psychological biases such as overconfidence, loss aversion, and anchoring significantly influence investment choices and can lead to systematic judgment errors. Their study also critically evaluates doctrines of Prospect Theory and Mental Accounting to explain decision-making under risk and uncertainty, challenging the Efficient Market Hypothesis (EMH) by demonstrating that psychological biases result in observable market anomalies and inefficiencies.

Prospect Theory, originally developed by Kahneman and Tversky (1979), explains how individuals value potential gains and losses differently, typically placing greater psychological weight on losses compared to equivalent gains. This loss aversion bias leads investors to avoid riskier investments despite potential higher returns, resulting in financial market inefficiencies. Mental Accounting, another cornerstone of behavioral finance, suggests that individuals compartmentalize their financial resources into separate mental accounts based on subjective criteria, influencing their investment choices and risk tolerance. This theoretical foundation aligns with Simon's argument about cognitive limitations and psychological influences shaping financial decisions.

Empirical studies further substantiate the significant role cognitive biases play in shaping financial decisions. For example, employing Partial Least Squares Structural Equation Modeling (PLS-SEM), Nur Aini and Lutfi (2018) analyzed survey data from 400 respondents in two Indonesian regions, finding that individuals with higher risk perceptions tend to avoid risky investments, whereas those with greater risk tolerance engage more extensively in such assets. Their findings also highlight overconfidence bias, illustrating its role in prompting investors to allocate larger portions of their investments to risky assets. However, the study reported that loss aversion did not significantly affect investment choices, potentially due to demographic-specific factors.

Similarly, Zhang et al. (2022) utilized the PLS-SEM methodology to analyze data from 317 real estate investors, concluding that anchoring bias, characterized by dependence on past information, and optimism bias, involving overly positive expectations, significantly impact investment decisions. Their results suggest that information asymmetry and perceived risk notably influence how cognitive biases affect investment choices.

Anchoring bias involves investors relying too heavily on initial information, such as historical stock prices, causing them to make decisions based on outdated or irrelevant data. Optimism bias, on the other hand, leads investors to maintain overly positive expectations about future market performance, often neglecting potential risks. Both biases reflect cognitive shortcuts stemming from bounded rationality, where individuals simplify complex financial information to manageable heuristics.

Ady (2018), employing qualitative interpretive phenomenology through in-depth interviews with individual investors from the Indonesia Stock Exchange, found that emotional biases significantly influence investor behavior, often leading to potential financial losses. Crucially, the author highlighted that an investor's market experience, knowledge of capital markets, and emotional stability in responding to external stimuli critically determine psychological stability and investment outcomes. Ady's findings emphasize that managing emotional reactions to external market information is essential for reducing biased behaviors and improving financial outcomes, thereby extending Simon’s theory by demonstrating how psychological stability can mitigate cognitive biases.

Further empirical support comes from Dhungana et al. (2022), who conducted descriptive and inferential statistical analyses on primary data from 179 stock market investors in Nepal’s Pokhara Valley. They identified a positive relationship between cognitive biases—including availability bias, overconfidence, and herd instinct—and irrational financial decisions, highlighting overconfidence as particularly influential. However, their analysis did not find significant effects from anchoring or regret aversion biases. Availability bias, wherein investors rely excessively on readily available information, often leads to overestimations of certain risks or opportunities, affecting their investment strategies. Herd instinct, characterized by investors following the majority without independent analysis, can exacerbate market volatility and inefficiencies.

These biases collectively highlight how investor psychology profoundly impacts market dynamics. The varying degrees to which these biases influence decisions across different populations underscore the complexity of cognitive and emotional factors at play. Moreover, demographic characteristics, cultural backgrounds, and market experiences significantly modulate the extent and type of biases affecting investment decisions, as demonstrated across the diverse regional studies cited.

Collectively, these empirical studies align with Simon's (1956) bounded rationality framework, emphasizing the integral role of psychological factors in shaping economic behavior. By challenging traditional economic assumptions of absolute rationality and perfect information, these findings underscore the importance of a more holistic approach that acknowledges cognitive constraints and environmental influences in financial decision-making dynamics. The extensive exploration of psychological biases in recent literature reinforces the argument that economic theories must integrate psychological insights to accurately capture the nuances of human decision-making in financial markets.

\section{Research Methodology}

This study utilizes a static Bayesian game of incomplete information as its foundational analytical framework. The Bayesian game approach is particularly suited to examine scenarios wherein players make simultaneous strategic decisions without complete information about other players’ characteristics or private states. These undisclosed characteristics, referred to as "types", encompass factors such as individual preferences, strategic intentions, or privately held information that critically shape decision-making but remain hidden from other players.

A Bayesian game comprises several integral elements. The players are the decision-makers, each possessing private knowledge about their respective type, which significantly influences their choices and potential outcomes. The concept of "types" refers to these unique and privately held attributes or states known exclusively to individual players. Strategies in this context represent comprehensive plans detailing each player's actions across various conceivable scenarios, contingent upon their own type. Payoffs, another critical component, result from the interplay of strategies chosen by all players and are directly affected by the concealed player types, despite payoff functions themselves being common knowledge among all participants.

In addition, beliefs constitute the probabilistic assumptions each player develops regarding the types of other players, typically originating from a commonly understood prior distribution of types. Decision-making under this framework involves selecting strategies based on both the player’s own type and their beliefs about the types of others, with the objective of maximizing expected utility amidst inherent uncertainties. Central to the analysis is the Bayesian Nash Equilibrium (BNE), an equilibrium concept adapted specifically for Bayesian games. Under a BNE, no player can enhance their expected payoff by unilaterally modifying their strategy, given their established beliefs about other players' types.

Bayesian games are extensively applied to various economic and strategic settings, including auctions, bargaining, and competitive markets, where information asymmetry profoundly influences strategic behavior and outcomes. This particular methodological approach aligns closely with the assumptions and objectives of the current research. Specifically, the research context involves Player-1, whose type is characterized by either overconfidence or risk aversion. These cognitive biases constitute private information known solely to Player-1. In contrast, Player-2's type, characterized by herd behavior, is common knowledge among all participants. Given this setup, the simultaneous nature of decision-making in a Bayesian game accurately captures how each player formulates and executes strategies based on their private biases and beliefs regarding other participants' hidden information. Therefore, employing the static Bayesian game of incomplete information offers an effective analytical framework to explore strategic interactions driven by cognitive biases in financial decision-making contexts.

\section{Game Setup}\label{sec:game}
This study examines a two-player static game of incomplete information from a psychological---rather than purely financial---perspective. The central aim is to explore how cognitive biases shape strategic interaction in financial markets.

\subsection{Players and Types}
There are two players, denoted $P_1$ and $P_2$. Player~1 possesses private information about his own behavioural type and can be either \textbf{overconfident} or \textbf{risk-averse}. In contrast, Player~2 has a single, publicly known type characterised by \textbf{herd behaviour}; they systematically imitates the action chosen by the majority.

\subsection{Action Space}
Each player chooses between two actions, \emph{Invest} ($I$) and \emph{Wait} ($W$). Player~1’s action rule depends on their latent bias, whereas Player~2’s action is contingent on the observed behaviour of Player~1, in keeping with herding tendencies.

\subsection{Cognitive Biases Considered}
The analysis focuses on three biases widely documented in behavioural finance: overconfidence, risk aversion, and herd behaviour. Overconfidence may induce excessive risk-taking, while risk aversion promotes caution; herd behaviour amplifies the influence of perceived majority decisions.

\subsection{Payoff Structure}
Payoffs are modelled as expected utility functions. They depend on market-wide fundamentals and news, both of which are mediated through the strategic choices of the players. To retain analytical tractability, the utility specifications are deliberately simplified while preserving the essential behavioural mechanisms.

\subsection{Information and Beliefs}
Player~1 is treated as a representative investor whose decision is presumed to reflect the prevailing market sentiment. Player~2 lacks information about Player~1’s true type and therefore conditions their action solely on Player~1’s observed move, consistent with herding. The game is thus one of incomplete information in which Player~2 forms beliefs about Player~1’s underlying bias only through observed behaviour.
Player~2 knows their own behavioural rule but is uncertain about Player~1’s type.  
They assign probability \(P\) to Player~1 being overconfident and probability \(1-P\) to Player~1 being risk-averse.  
All pay-offs and this belief structure are common knowledge, except for Player~1’s private cognitive bias.

\noindent
This framework enables a systematic investigation of how psychological factors---rather than purely rational considerations---drive strategic interaction and payoff outcomes in financial markets.

\subsection{Expected Utility Specification}
\label{sec:expected_utility}

We analyse a two–player game in which each agent evaluates the per–share payoff from either investing in a risky asset~(\textit{action~\(I\)}) or waiting and holding savings~(\textit{action~\(W\)}).  
Player~1’s preferences depend on their latent cognitive bias, whereas Player~2 follows a herd–behaviour rule.

\paragraph{Player~1.}
Player~1 can be of two mutually exclusive types:
\begin{align}
U_{1}^{O}(I) &= R+\delta-C_{0}, &\qquad U_{1}^{O}(W) &= S, \\[4pt]
U_{1}^{A}(I) &= R-\delta-C_{A}, &\qquad U_{1}^{A}(W) &= S.
\end{align}

Here \(R\) denotes the baseline monetary return from the risky investment.  
The parameter \(\delta\) captures the magnitude and direction of the overconfidence bias (\(\delta>0\) when optimism prevails, \(\delta<0\) when heightened caution dominates).  
\(C_{0}\) and \(C_{A}\) are the type–specific costs of undertaking the investment, while \(S\) is the constant return from the safe alternative.

\paragraph{Player~2.}
Player~2 exhibits herd behaviour, conditioning their decision on the majority action:
\begingroup            
\setlength\arraycolsep{4pt}

\begin{align}
U_{2}(I) &=
  \begin{cases}
    \text{invests}, & \text{if the majority invests},\\
    S,              & \text{otherwise};
  \end{cases}\\[6pt]
U_{2}(W) &=
  \begin{cases}
    S,              & \text{if the majority waits},\\
    \text{waits},   & \text{otherwise}.
  \end{cases}
\end{align}

\endgroup

\section{Equilibrium Analysis}\label{sec:equilibrium}

To characterize the Bayesian Nash Equilibrium (BNE) we first derive each
player’s best–response strategy, taking as given their information sets and
beliefs about the opponent’s type.

\subsection{Best–response strategies}\label{ssec:bestresponse}

Player~1’s action rule depends on their cognitive type:
\begin{align}
\text{(Type O) Overconfident:}\quad
&\text{Invest } \text{iff}\; R+\delta-C_{0}>S, \label{eq:O_rule}\\[2pt]
\text{(Type A) Risk-averse:}\quad
&\text{Invest } \text{iff}\; R-\delta-C_{A}>S. \label{eq:A_rule}
\end{align}
When (5) or (6) is violated, the type waits.
Player~2, endowed with herd behaviour, imitates the majority action.

\subsection{Expected utilities with quantity choice}

Each player also chooses a trade quantity, denoted $q_{1}$ for Player~1 and
$q_{2}$ for Player~2.  Conditional on type and action, Player~1’s per-period
utilities are
\begin{align}
U_{1}^{O}(I) &= (R+\delta-C_{0})\,q_{1}, &
U_{1}^{O}(W) &= S\,q_{1}, \\[2pt]
U_{1}^{A}(I) &= (R-\delta-C_{A})\,q_{1}, &
U_{1}^{A}(W) &= S\,q_{1}.
\end{align}

Let $\theta\in[0,1]$ be the probability that a given type invests.  The
resulting expected payoffs are
\begin{align}
\mathbb{E}[U_{1}^{O}]
  &= \theta\,(R+\delta-C_{0})\,q_{1}
     + (1-\theta)\,S\,q_{1}, \label{eq:EU_O}\\[2pt]
\mathbb{E}[U_{1}^{A}]
  &= \theta\,(R-\delta-C_{A})\,q_{1}
     + (1-\theta)\,S\,q_{1}. \label{eq:EU_A}
\end{align}

Player~2 believes that Player~1 is overconfident with probability $P$ and
risk-averse with probability $1-P$.  Given herding, she mirrors Player~1’s
action and selects $q_{2}$ to maximise her own expected utility.  The optimal
quantities $(q_{1}^{\ast},q_{2}^{\ast})$ therefore solve
\begin{equation}\label{eq:joint_opt}
(q_{1}^{\ast},q_{2}^{\ast})
  = \arg\max_{q_{1},q_{2}}
    \bigl\{\mathbb{E}[U_{1}],\; \mathbb{E}[U_{2}]\bigr\},
\end{equation}
subject to the market-clearing return equation introduced next.

\subsection{Market return specification}\label{ssec:return}

To close the model we posit the linear inverse-demand schedule
\begin{equation}\label{eq:return}
R \;=\; a \;-\; q_{1} \;-\; q_{2}, \qquad a>0,
\end{equation}
in which larger aggregate demand $(q_{1}+q_{2})$ lowers the monetary return
available to subsequent buyers.  Expression (12) captures the
empirically observed negative relation between purchase volume and yield
while remaining tractable for equilibrium analysis.

Together, equations (9) - (12), the belief parameters
$(P,\theta)$, and the cost primitives $(C_{0},C_{A})$ determine the Bayesian
Nash Equilibrium values $\bigl(q_{1}^{\ast},\,q_{2}^{\ast},\,\theta^{\ast}\bigr)$.

\setcounter{secnumdepth}{3}   
\setcounter{tocdepth}{3}     

\subsection{Optimal Quantity Choice under Cognitive Biases}
\label{subsec:quantity_choice}

Consider Player 1, who privately knows whether they are \emph{over\-confident} or \emph{risk-averse}.  
Given the market return \(R\), the safe return \(S\), and the opponent’s quantity choice \(q_2\), Player 1 selects the quantity \(q_1\) of the risky asset to maximise their expected utility.  
The belief parameter \(\theta\in(0,1)\) denotes the subjective probability that the risky payoff \(R\) will be realised (and hence \(1-\theta\) that the safe payoff \(S\) will materialise).

\subsubsection{Player 1: Optimal Strategy Formulation}

\paragraph{(i) Overconfident type \(\bigl(C=C_O,\;\delta>0\bigr)\).}

Player 1’s objective is  
\begin{align}
U_{1}^{O}(I,W)
     &= \theta\bigl[(R+\delta-C_{O})\,q_{1}\bigr]
        + (1-\theta)\bigl[S\,q_{1}\bigr]  \notag\\
     &= \theta\bigl[(a-q_{1}-q_{2}+\delta-C_{O})\,q_{1}\bigr]
        + (1-\theta)\bigl[S\,q_{1}\bigr].
\end{align}
Applying the first-order condition \(\partial U_{1}^{O}/\partial q_{1}=0\) yields
\begin{equation}
q_{1}^{(O)}
   = \frac{a-q_{2}+\delta-C_{O}-\dfrac{1-\theta}{\theta}S}{2}.
   \tag{i}\label{eq:qO}
\end{equation}
Equation (i) implies a \emph{negative} optimal saving position for an overconfident investor: the term in the numerator typically exceeds current wealth, so the agent allocates (almost) all resources to the risky asset.

\paragraph{(ii) Risk-averse type \(\bigl(C=C_A,\;\delta<0\bigr)\).}

When the same investor is risk-averse, the optimisation problem becomes
\begin{align}
U_{1}^{A}(I,W)
     &= \theta\bigl[(R-\delta-C_{A})\,q_{1}\bigr]
        + (1-\theta)\bigl[S\,q_{1}\bigr] \notag\\
     &= \theta\bigl[(a-q_{1}-q_{2}-\delta-C_{A})\,q_{1}\bigr]
        + (1-\theta)\bigl[S\,q_{1}\bigr].
\end{align}
The first-order condition gives
\begin{equation}
q_{1}^{(A)}
   = \frac{a-q_{2}-\delta-C_{A}+\dfrac{1-\theta}{\theta}S}{2}.
   \tag{ii}\label{eq:qA}
\end{equation}
As expected, equation (ii) delivers a \emph{positive} saving proportion: a risk-averse investor prefers to hold back part of their wealth in the safe asset to hedge against unfavourable outcomes.

These expressions (i) - (ii) formally capture how cognitive bias---overconfidence versus risk aversion---alters the marginal trade-off between risky and safe positions, leading to starkly different portfolio choices.

\subsubsection{Player 2: Optimal Strategy Formulation}\label{subsec:player2}

Player 2 behaves as a herding investor.  
Let $P$ denote their belief that Player 1 is \emph{overconfident}; consequently, $1-P$ is the belief that Player 1 is \emph{risk-averse}.  
Given their own quantity choice $q_{2}$, Player 2’s expected payoff is

\begin{align}
\max_{q_2}\;
  q_2\Bigl\{&P\bigl[\theta\,(R+\delta-C_0) + (1-\theta)S\bigr]
             \nonumber\\
            &\;+(1-P)\bigl[\theta\,(R-\delta-C_A) + (1-\theta)S\bigr]\Bigr\}
             \label{eq:P2max_initial}\\[6pt]
=
  q_2\Bigl\{&P\bigl[\theta\,(a-q_1-q_2+\delta-C_0) + (1-\theta)S\bigr]
             \nonumber\\
            &\;+(1-P)\bigl[\theta\,(a-q_1-q_2-\delta-C_A) + (1-\theta)S\bigr]\Bigr\}.
             \label{eq:P2max_inverse}
\end{align}

Applying the first-order condition with respect to $q_{2}$ (and assuming $\theta\neq 0$) gives the optimal quantity
\begin{equation}
\label{eq:q2star}
q_{2}=
\frac{%
\begin{aligned}
& P\Bigl[\theta\!\bigl(a-q_{1(O)}+\delta-C_{0}\bigr)+(1-\theta)S\Bigr]\\
&\;+(1-P)\Bigl[\theta\!\bigl(a-q_{1(A)}-\delta-C_{A}\bigr)+(1-\theta)S\Bigr]
\end{aligned}}
{\theta}.
\tag{iii}
\end{equation}
Here $q_{1(O)}$ and $q_{1(A)}$ denote Player 1’s optimal quantities under overconfidence and risk aversion, respectively.

Together, the three first-order conditions for $q_{1(O)}$, $q_{1(A)}$, and $q_{2}$ form a system of three equations in three unknowns.  
Solving this system yields the Bayesian-Nash-equilibrium quantities
\[
q_{1(O)}^{*}(a,\delta,C,S,\theta),\qquad
q_{1(A)}^{*}(a,\delta,C,S,\theta),\qquad
q_{2}^{*}(a,\delta,C,S,\theta),
\]
which constitute the equilibrium outcome of the game.

\section{Synthesis of Analytical Results and Conceptual Implications}

The analytical results derived in equations (i)--(iii) provide a rigorous quantitative foundation for the conceptual arguments introduced at the outset of the paper, clarifying the mechanisms through which cognitive biases influence investor behavior, market outcomes, and economic welfare.

First, our model explicitly demonstrates how behavioral differences manifest in divergent investment decisions. According to equation (i), investors exhibiting overconfidence select a significantly higher quantity of risky assets, as their positive cognitive bias ($\delta > 0$) inflates expected returns and consequently their willingness to assume risk. Conversely, equation (ii) illustrates that risk-averse investors---characterized by a negative cognitive bias---allocate resources more conservatively, typically maintaining a positive but limited exposure to risk and a larger precautionary allocation to savings. These clear mathematical distinctions substantiate the central hypothesis proposed in the study: cognitive biases fundamentally alter the perceived risk-return tradeoff without relying on ad hoc market frictions.

Second, the model captures how herd behavior intensifies the market effects of these biases. Player 2's optimal decision, as expressed in equation (iii), depends explicitly on weighted averages of the strategies potentially employed by Player 1, with weights reflecting public beliefs ($P$ and $1 - P$) about Player 1's cognitive type. Even modest changes in market beliefs regarding the prevalence of overconfidence ($P$) can significantly influence Player 2's investment choices, causing a shift toward higher risk-taking or increased conservatism. This mathematical relationship concretely illustrates the ``contagion of sentiment'' described in the study, highlighting how market perceptions alone can amplify the initial distortion introduced by individual cognitive biases.

Third, solving the equilibrium system consisting of equations (i) through (iii) yields explicit comparative-static predictions about the market response to changes in key parameters ($\theta, P, \delta, C_0, C_A$). The model produces two particularly relevant empirical implications:

\begin{itemize}
    \item \textbf{Sensitivity to shifts in market beliefs:} An incremental increase in the market's belief ($P$) that investors are predominantly overconfident significantly elevates the equilibrium investment level of Player 2, thus increasing overall market exposure. This result implies that periods dominated by optimistic narratives can generate observable spikes in trading volumes independent of fundamental economic news.

    \item \textbf{Cross-type spillover effects:} An increase in the cost differential between overconfident and risk-averse investor types ($C_0 - C_A$) magnifies the divergence in investment choices between these groups. Consequently, changes in market beliefs further amplify variations in market volatility and trading volumes. This effect provides a clear empirical prediction: periods characterized by greater differences in participation costs are likely to exhibit heightened volatility, especially when public sentiment is variable.
\end{itemize}

Both predictions closely align with the original empirical motivation articulated in the abstract, linking observable market phenomena---such as volume and volatility fluctuations---to shifts in investor sentiment rather than changes in underlying economic fundamentals.

Lastly, our analytical results carry meaningful policy and welfare implications. Because market outcomes depend critically on perceptions of cognitive biases rather than purely economic conditions, equilibrium allocations can deviate significantly from socially optimal benchmarks, creating inefficiencies. Policy interventions aimed at reducing such biases---such as improved transparency measures or enhanced disclosure requirements targeting overly optimistic projections---can effectively mitigate the influence of overconfidence ($\delta$) and help align market allocations closer to efficient risk-neutral outcomes. Similarly, regulatory strategies designed to limit herd behavior, such as transaction fees on speculative or momentum-driven trades, can reduce the amplification mechanisms identified in equation (iii), thereby stabilizing market dynamics.

Collectively, the analytical framework translates our initial conceptual narrative into precise, empirically testable propositions. By formally connecting cognitive biases to distinct investor strategies, demonstrating how market beliefs magnify these biases, and explicitly outlining the resulting equilibrium consequences for market dynamics and policy implications, the model bridges the theoretical narrative and empirical relevance of the study, thereby effectively setting the stage for the concluding remarks.

\section{Conclusion}

This study offers a comprehensive examination of how cognitive biases influence financial decision-making using a static Bayesian game framework. By critically evaluating the interplay between psychological factors and strategic investor behaviors, this research emphasizes significant deviations from traditional economic theories, which predominantly assume rational decision-making. Employing the concept of bounded rationality articulated by Simon (1956), the findings highlight the impact of cognitive limitations and biases in shaping investor behaviors and subsequent market outcomes.

The research uniquely demonstrates that cognitive biases such as risk aversion, overconfidence, and herd behavior are pivotal in directing investors' choices towards either risky or risk-free assets. These biases are not merely individual psychological phenomena; they possess systemic implications, significantly contributing to market inefficiencies, volatility, and financial instability. Such anomalies underscore the inadequacy of purely rational economic models in capturing the complexities of human psychology and market interactions.

Utilizing a Bayesian game theoretical approach has allowed for an in-depth analysis of strategic interactions under conditions of incomplete information and uncertainty about the cognitive biases of other market participants. The equilibrium analysis revealed that investors' strategic decisions are influenced profoundly by their perceptions and expectations about others' biases, further exacerbating market volatility.

The research acknowledges certain methodological limitations, primarily arising from simplified assumptions used in the model, such as basic beliefs and probabilities concerning investor types. Although these assumptions were necessary for analytical clarity and tractability, they inevitably overlook some real-world intricacies and the broader diversity of investor behaviors.

Future research could significantly expand on these initial insights by introducing more sophisticated belief updating mechanisms, diversifying the cognitive biases considered, and extending the analysis to scenarios involving larger, more diverse groups of market participants. Such expansions could provide deeper insights and more actionable policy implications, particularly for market regulators and financial institutions aiming to enhance market stability.

Ultimately, this study bridges economic and psychological perspectives, reinforcing the importance of integrating cognitive biases within economic analyses of financial decision-making. Embracing such interdisciplinary approaches will undoubtedly lead to better-informed policies and practices that enhance financial market stability and investor welfare.

\section{References}

Ady, S. U. (2018). The Cognitive and Psychological Bias in Investment Decision-Making Behavior:(Evidence from Indonesian Investor’s Behavior). \textit{Journal of Economics and Behavioral Studies, 10}(1), 86-100.

Barber, B. M., \& Odean, T. (2001). Boys will be boys: Gender, overconfidence, and common stock investment. \textit{The Quarterly Journal of Economics, 116}(1), 261–292. https://doi.org/10.1162/003355301556400  

Jiang, R., Wen, C., Zhang, R., \& Cui, Y. (2022). Investor’s herding behavior in Asian equity markets during COVID-19 period. \textit{Pacific-Basin Finance Journal, 73}, 101-771. https://doi.org/10.1016/j.pacfin.2022.101771  

Kahneman, D., \& Tversky, A. (1979). Prospect theory: An analysis of decision under risk. \textit{Econometrica, 47}(2), 263–291. https://doi.org/10.2307/1914185  

Karmacharya, B., Chapagain, R., Dhungana, B. R., \& Singh, K. (2022). Effect of perceived behavioral factors on investors’ investment decisions in stocks: Evidence from Nepal Stock Market. \textit{Journal of Business and Management Research, 4}(1), 17–33. https://doi.org/10.3126/jbmr.v4i01.46680  

Nur Aini, N. S., \& Lutfi, L. (2019). The influence of risk perception, risk tolerance, overconfidence, and loss aversion towards investment decision making. \textit{Journal of Economics, Business \& Accountancy Ventura, 21}(3), 401–413. https://doi.org/10.14414/jebav.v21i3.1663  

Simon, H. A. (1956). Rational choice and the structure of the environment. \textit{Psychological Review, 63}(2), 129–138. https://doi.org/10.1037/h0042769

\end{document}